# Unexpected Reversed Piezoelectric Response in Elemental Sb and Bi Monolayers

Yunfei Hong[1], Junkai Deng[1,*], Qi Kong[1], Xiangdong Ding[1], Jun Sun[1], and Jefferson Zhe Liu[2,*]

[1]*State Key Laboratory for Mechanical Behavior of Materials, Xi'an Jiaotong University, Xi'an 710049, China*

[2]*Department of Mechanical Engineering, The University of Melbourne, Parkville, VIC 3010, Australia*

[*]Corresponding author

E-mail: junkai.deng@mail.xjtu.edu.cn; zhe.liu@unimelb.edu.au

## ABSTRACT

Sb and Bi monolayers, as single-elemental ferroelectric materials with similar atomic structure, hold intrinsic piezoelectricity theoretically, which makes them highly promising for applications in functional nano-devices such as sensors and actuators. Here, using first-principles calculations, we systematically explore the piezoelectric response of Sb and Bi monolayers. Our findings reveal that Sb exhibits a negative piezoelectric response, whereas Bi displays a positive one. This discrepancy is attributed to the dominant role of different atomic internal distortions (internal-strain terms) in response to applied strain. Further electron-density distribution analysis reveals that the atomic bonding in Sb tends to be covalent, while the atomic bonding in Bi leans more towards ionic. Compared to the Sb monolayer, the Bi monolayer is distinguished by its more pronounced lone-pair orbitals electrons and associated larger Born effective charges. The Coulomb repulsions between lone-pair orbitals electrons and the chemical bonds lead to the Bi monolayer possessing more prominent atomic folds and, consequently, more significant atomic distortion in the *z*-direction under strain. These differences result in a considerable difference in internal-strain terms, ultimately leading to the reversed piezoelectric response between Sb and Bi monolayers. The present work provides valuable insights into the piezoelectric mechanism of 2D ferroelectric materials and their potential applications in nano-electronic devices.

**INTRODUCTION**

Piezoelectric materials, due to their ability to be electrically polarized under an externally applied strain and, conversely, to be deformed by an applied voltage, provide an effective means of transducing signals between mechanical strain and electrical polarization [1]. The piezoelectric performance of materials is typically measured by the piezoelectric coefficient [2,3]. In most cases, the piezoelectric coefficient is positive, indicating that polarization is more likely to increase during stretching [4,5]. However, in recent years, researchers have discovered some exceptions where certain materials exhibit a negative piezoelectric response. For instance, researchers have observed negative longitudinal piezoelectric effects in the ferroelectric polymer polyvinylidene fluoride (PVDF) and its copolymers [6–9]. Moreover, theoretical predictions suggest the presence of significant negative piezoelectric responses in some specific organic molecular ferroelectrics [10], hexagonal *ABC* ferroelectrics [11], layered van der Walls ferroelectrics [12–14], group IV-V $MX_2$ [15,16], $HfO_2$ ferroelectric [17,18], and sphalerite compounds [19–21].

The piezoelectric effect arises from the influence of strain on the redistribution of electric charges, involving ion displacement and the changes in Born effective charges. As a result, the total piezoelectric coefficient ($e_{22}$) can be decomposed into the "clamped-ion" term ($e_{22}^0$) and the "internal-strain" term ($e_{22}'$) [22]. Recently, Yubo Qi and Andrew M. Rappe discovered the "lag of Wannier center" effect, resulting from the decay of the coulomb repulsion between electrons with the separation of anions and cations [13]. Therefore, the Wannier centers generally fail to follow the anions (in fractional coordinates) completely upon a tensile strain, contributing to the consistently negative clamped-ion term ($e_{22}^0$) in piezoelectric materials. Besides, Liu and Cohen revealed that the negative piezoelectric effect originates from the dominance of the negative clamped-ion term ($e_{22}^0$) over the internal-strain term ($e_{22}'$) [11]. Therefore, the sign (positive or negative) of the piezoelectric coefficient depends on the competition between those two terms.

Piezoelectric materials have widespread applications in electromechanical systems and electronic devices such as energy harvesters, pressure sensors, actuators, strain-tuned electronics, and optoelectronics [1,23,24]. With the development of innovative miniaturized electronic devices, 2D piezoelectric materials have drawn great interest [25,26]. So far, many 2D piezoelectric materials have been reported [5,27–30]. Remarkably, Sb and Bi ferroelectric monolayers have recently attracted significant attention due to their unique single-element attribute [31,32]. In 2023, a study further confirmed the ferroelectricity of Bi monolayer experimentally [33]. Due to their ferroelectric nature, Sb and Bi monolayers must have intrinsic piezoelectricity. Furthermore, unlike traditional compound piezoelectric materials, where the polarization is primarily contributed by ions, the polarization in Sb and Bi monolayers arises solely from the electron contribution, resulting in the piezoelectric response exclusively determined by the electrons redistribution under applied strains. As the electron redistribution behavior is highly uncertain and closely related to the elemental species, crystal structure, and bonding types [34,35], studying the piezoelectric response of Sb and Bi monolayers is of great significance for exploring their electromechanical coupling properties, elucidating their physical mechanisms.

In this paper, we investigate the piezoelectric response of Sb and Bi monolayers. Our results demonstrate that Sb exhibits a negative piezoelectric response while Bi exhibits a positive one. By analyzing the piezoelectric coefficient in terms of clamping-ion ($e_{22}^0$) and internal-strain ($e_{22}'$) terms contributions, we find that the origin of the difference is related to the internal-strain term. Furthermore, we explain the physical mechanism of the polarization change during the internal strain process using the density of states (DOS), Born effective charges, and atomic displacement calculations. This work can provide deep insight into the mechanism of piezoelectricity and can be used to design potential devices employing the anomalous electromechanical response.

**METHODS**

Based on the density functional theory (DFT), first-principles calculation has been performed using the projected augmented plane-wave method as implemented in the Vienna Ab-initio Simulation Package (VASP) [36–39]. Perdew-Burke-Ernzerhof (PBE) functional of generalized gradient approximation [40] was used to treat electron exchange and approximation with valence configurations $5s^25p^3$ for Sb and $6s^26p^35d^{10}$ for Bi. The periodic boundary condition was applied along *x* and *y* axis and vacuum space about 20 Å was applied along *z* axis to safely avoid artificial interaction. The energy cutoff was set to be 500 eV. The Brillouin zone was sampled in *k* space within 15×15×1 mesh for Sb and Bi monolayers. The positions of all atoms were fully relaxed until the energy between two consecutive steps was less than $10^{-6}$ eV, and the maximum force was 0.001 eV/Å on each atom. The spin-orbit coupling (SOC) was considered in all calculations [41–43]. Poisson's ratio effect [44–46] was considered in the process of applying uniaxial strain and the electric polarization was computed using the Berry-phase method [47,48].

**RESULTS AND DISCUSSION**

Sb and Bi monolayers are crystallized in the asymmetric washboard structures with space group *Pmn2$_1$* and four atoms per unit cell. The Sb and Bi monolayers have been predicted as single-elemental ferroelectric materials, which was later experimentally confirmed for the Bi monolayer in 2023 [33]. In contrast to black phosphorus, which has only one chemical bonding state, the asymmetric washboard structure of Sb and Bi monolayers allows for different chemical surroundings, enabling the adoption of two bonding states in these elemental materials.

The optimized lattice constants of Sb and Bi monolayers are *a*=4.362 Å, *b*=4.736 Å and *a*=4.581 Å, *b*=4.913 Å, respectively (FIG 1). The Sb and Bi monolayers possessed a stable structure as demonstrated by the phonon spectrum (FIG S1). Our DFT calculations revealed that 2D Sb is a direct band gap semiconductor (with a band gap of 0.22 eV), while 2D Bi is an indirect band gap semiconductor (with a band gap of 0.26 eV). The energy band structure showed that the atomic buckling of Sb and Bi monolayers splits the energy level of two $p_z$ orbitals, resulting in two states in one element. In the case of the Sb monolayer (FIG 1c), the conduction band and valence band at the $\Gamma$ point are mainly contributed by $p_z$ orbitals of $Sb^+$ and $Sb^-$ atoms, respectively. The $p_z$ orbitals of $Sb^-$ are occupied and the $p_z$ orbitals of $Sb^+$ are empty, indicating the electron transfer from $Sb^+$ to $Sb^-$. The unit cell of the Sb monolayer contains two sub-layers and each sub-layer has one $Sb^-$ atom and one $Sb^+$ atom. Similarly, electron transfer also occurs in Bi monolayer (FIG 1d).

The results of Berry phase analysis revealed that the spontaneous polarizations along the *y* direction of Sb and Bi monolayers were determined as -38 pc/m and -43 pc/m, respectively. Besides, it is noted that the *z* coordinates of $Sb^+(Bi^+)$ and $Sb^-(Bi^-)$ atoms are significantly different, leading to the considerable tilting of Sb-Sb bonding within the sub-layer. This is different from non-polarized black phosphorene (space group *Pmna*), where the *z* coordinates of the P atoms within a sub-layer are identical [49]. Thus, when the Sb and Bi monolayers (with the space group of *Pmn2₁*) were transformed to a "distorted" structure with the space group of *Pmna*, the same as the Phosphorene, eliminating the difference in *z* coordinates could give rise to the non-polarization state. It is confirmed by removing the energy level splitting of the two $p_z$ orbitals in such a distorted structure (FIG S2), indicating the chemical bonding state only. Thus, the spontaneous polarization of Sb (Bi) was highly dependent on the *z* coordinates of atoms.

To explore the piezoelectric response of Sb and Bi monolayers, the polarization changes with uniaxial strain (-0.5% ~ 0.5%) in the polarization direction (*y*) were determined as shown in Fig 2a and 2b, respectively. The polarization response to fully relaxed Sb and Bi monolayers under different strains were approximately fitted into straight lines, and their slopes were labeled as the piezoelectric tensor elements ($e_{22}$) for full structural relaxation. As reported in TABLE I, the $e_{22}$ were found to be negative (-1.971×10⁻¹⁰ c/m) for Sb and positive (7.653×10⁻¹⁰ c/m) for Bi, indicating that Sb and Bi monolayers exhibit completely reversed piezoelectric responses, despite having a similar structure. We double confirmed the reversed piezoelectric responses with the strain of -5% to 5% (FIG S3).

In modern piezoelectric theory, the total piezoelectric coefficient ($e_{22}$) can be decomposed into the "clamped-ion" term ($e_{22}^0$) and the "internal-strain" term ($e'_{22}$) [22].

$$e_{22} = e_{22}^0 + e'_{22} \tag{1}$$

The $e_{22}^0$ value was computed with the internal atomic fractional coordinates fixed at their zero-strain states, corresponding to the uniform scaling of the atomic positions in the unit cell as the strain state is changed. This "clamped-ion" piezoelectric coefficient features the change of Born effective charges, reflecting the redistribution of electrons with respect to a homogeneous strain.

The $e'_{22}$ is an internal-strain term arising from internal microscopic polar distortion in response to a microscopic strain applied in the *y* direction (Eq 2).

$$e'_{22} = \sum_s \frac{e}{V}\left(Z^*_{12}\frac{a\times\partial\mu_1(s)}{\partial\eta_2} + Z^*_{22}\frac{b\times\partial\mu_2(s)}{\partial\eta_2} + Z^*_{32}\frac{c\times\partial\mu_3(s)}{\partial\eta_2}\right) \tag{2}$$

Here, $\mu_1, \mu_2, and\ \mu_3$ represent the fractional atomic coordinates, while $\eta_2$ denotes the macroscopic strain applied in the *y* direction. The variable *s* runs over the atoms in the unit cell of Sb/Bi, and *V* represents the volume (in 3D systems) or area (in 2D systems) of the unit cell. The parameters *e*, *a*, *b,* and *c* correspond to the electron charge and the lattice constants associated with the atomic coordinates, respectively. The Born effective charges related to the displacement of $\mu_1, \mu_2, and\ \mu_3$ are represented by $Z^*_{12}, Z^*_{22}, and\ Z^*_{32}$, respectively. Unlike the clamped-ion term, the internal-strain term mainly reflects the polarization contribution of internal atomic distortion, even though the Born effective charges vary slightly. Due to the change of Born effective charges, the internal-strain term, which is difficult to calculate directly, is usually obtained by subtracting the clamped-ion term from the total piezoelectric coefficient.

The values of the clamped-ion ($e_{22}^0$) and internal-strain ($e'_{22}$) terms are summarized in TABLE I. The clamped-ion term values for Sb (-2.623×10$^{-10}$ c/m) and Bi (-2.157×10$^{-10}$ c/m) are both negative and similar magnitude. The internal-strain term values for Sb and Bi are both positive, with Sb at 0.652×10$^{-10}$ C/m and Bi at 9.810×10$^{-10}$ C/m. It is notable that the internal-strain term value for Bi is over ten times larger than that of Sb. Consequently, the piezoelectric effect of the Sb monolayer is negative, as the negative clamped-ion piezoelectric response dominates over the internal-strain contribution. On the other hand, the piezoelectric response of Bi is dominated by the internal-strain term, resulting in a positive piezoelectric effect.

Due to the "lag of Wannier centers" effect, the clamped-ion term of piezoelectric response is always negative [13]. In the present work, the $e_{22}^0$ values of Sb and Bi monolayers are very close which may be attributed to their similar structures. It also indicates that the clamped-ion term is not responsible for the reversed piezoelectric coefficients of Sb and Bi, but rather the internal-strain term is.

The electric polarization ($P = \frac{\Sigma p}{V}$) is defined as the vector sum of electric dipole moments ($p$) per unit volume, and the Born effective charges and atomic displacements play important roles in polarization and the internal-strain term of piezoelectric coefficient (according to Eq 2). Therefore, the Born effective charges of Sb and Bi are calculated and summarized in TABLE II. The displacement of atoms in the $x$, $y$, and $z$ directions can change the electric dipole moment. However, the magnitude of the Born effective charges suggests that the displacement in the $y$ and $z$ directions may play a major role in the change of polarization, with the effect of displacement in the $z$ direction being more significant than in the $y$ direction (TABLE II). Due to the symmetry breaking, the Born effective charges ($Z_{32}^*$) of two atoms in the same state are reversed. The absolute values of $Z_{32}^*$ for Sb$^-$ (0.29$e$) and Sb$^+$ (0.84$e$) are quite different, indicating the transfer of electrons between atoms. A similar behavior was also observed in Bi monolayer.

In addition, the difference between $Z_{32}^*$ for positive and negative charged atoms within Bi monolayer (0.07$e$ and 3.28$e$) is even more significant than that of Sb monolayer, indicating that more electrons transfer from Bi$^+$ to Bi$^-$. More importantly, the Born effective charges $Z_{22}^*$ of Bi are more than twice as large as those of Sb, and the $Z_{32}^*$ of Bi$^+$ is almost four times as large as Sb$^+$, suggesting that the polarization of 2D Bi would change more dramatically at the same atomic displacement.

According to Eq 2, the atomic displacement ($\mu_i$) under uniaxial strain also plays an important role in the internal-strain term. Therefore, the structural parameters under uniaxial strain were measured, as shown in FIG 3. Both the Born effective charges and atomic displacements in the $x$ direction are almost zero and were not considered in this work. When the strain was applied in the $y$ direction, the lattice constant $b$ can be decomposed into $y_1$ and $y_2$ (FIG 3a and 3b). FIG 3c and 3d shows that the $y_2$ component always changed faster than $y_1$ in Sb or Bi monolayer. This suggests that rotating chemical bonds between sub-layers is easier than extending chemical bonds within a sub-layer. However, during the first clamped-ion process, the $y_1$ component constantly changed faster than $y_2$, resulting from the intrinsic length relationship ($y_1 \approx 5y_2$). Therefore, during the second internal-strain process (FIG 3e and 3f), the $y_1$ decreased while the $y_2$ increased, with the atomic displacements of Sb being about two times larger than that of Bi. However, the Born effective charges $Z^*_{22}$ of Sb are about half as large as those of Bi. It suggests that the products of $Z^*_{22}$ and atomic displacement of Sb and Bi have similar values, indicating the $y$-direction displacement is not responsible for the huge difference in the internal-strain terms ($e'_{22}$).

Furthermore, the structural parameter $z_1$ also decreased in both Sb and Bi when the stretching was applied in the $y$ direction (FIG 3c and 3d). During the first clamped-ion process, the $z_1$ is constant and equal to the value of the ground state. The displacement in the $z$ direction only occurred during the second internal-strain process. As shown in FIG 3e and 3f, the $z_1$ of Bi always changed faster than that of Sb. Under the same strain, the atomic displacement of Bi is about four times larger than that of Sb. The Born effective charges $Z^*_{32}$ of Bi are also about four times larger than those of Sb, making the product of $Z^*_{32}$ and displacements of Bi over ten times larger than that of Sb. As a result, the internal-strain term ($e'_{22}$) of Bi is one order of magnitude larger than that of Sb. It is demonstrated that the $z$-direction displacement and $Z^*_{32}$ play a dominant role in the internal-strain term of Sb and Bi monolayers.

To further explore the mechanism of Born effective charges differences between Sb and Bi monolayers, the density of states (DOS) was calculated to analyze their electronic structure (FIG 4). The first peak of

$Sb^-$ ($Bi^-$) near the valence band maximum (VBM) is higher than that of $Sb^+$ ($Bi^+$) atom, as highlighted in the figure by red boxes, indicating the electron transfer from $Sb^+$ ($Bi^+$) to $Sb^-$ ($Bi^-$). The electron-density distributions in the energy range of the first peak (-0.9 eV~0 eV for Sb and -0.35 eV~0 eV for Bi) exhibit distinct differences between Sb and Bi. In particular, the electron-density distribution of Bi is more localized than that of Sb. When comparing the electron-density distributions in the energy range of the second peak (-1.4 eV to -0.9 eV for Sb and -1 eV to -0.35 eV for Bi), there is no significant difference (FIG S4). It can be concluded that the electron-density distribution of the first peak primarily reflects the difference in bonding between the two materials.

The first peak of the Bi monolayer is narrower and higher than that of Sb, indicating the electron-density distribution is more localized, which is also reflected in the charge distribution plot, as shown in the comparison of bottom sub-figures in FIG 4. The electrons of Bi are predominantly localized within each single atom and appear like lone-pair orbitals, displaying the characteristics of ionic bonding. In contrast, some electrons of Sb are shared among neighboring atoms, exhibiting the characteristics of covalent bonds. Consequently, the Born effective charges of Bi are larger than those of Sb, resulting in the reversed piezoelectric response ultimately. The results indicate that negative piezoelectric effect is more likely to occur in covalent piezoelectrics which was also concluded in wurtzite piezoelectric semiconductors [50].

Furthermore, the pronounced lone-pair orbitals electrons around $Bi^-$ atoms (electron localization function (ELF) analysis shown in FIG S5) may be related to the intrinsic folds ($z_1$) of Bi. The Coulomb repulsion between the lone-pair orbitals electrons and the surrounding chemical bonds drives the $Bi^-$ away from $Bi^+$, creating the large intrinsic folds ($z_1$) of Bi [44,51]. The large intrinsic fold allows Bi atoms to undergo more significant displacements along the *z*-axis under uniaxial strain.

**CONCLUSION**

In summary, we investigated the piezoelectric response of Sb and Bi and found that Sb has a negative piezoelectric coefficient ($e_{22}$), while Bi has a positive one. It was discovered that Sb and Bi have similar clamped-ion terms $e_{22}^0$, but their internal-strain terms $e_{22}'$ differ by more than ten times, which plays the dominant role in their reversed piezoelectric response. In-depth electronic structure analysis revealed that the chemical bond of Bi is predominantly ionic, whereas the chemical bond of Sb is more covalent. This difference results in Bi having larger Born effective charges than Sb. Furthermore, the lone-pair electrons of Bi create more prominent folds and $z$-direction displacements. Consequently, the positive internal-strain term of Bi is much larger than that of Sb, dominating the piezoelectric coefficient. In contrast, the piezoelectric coefficient of Sb is dominated by the negative clamped-ion term. This work provides valuable insights into the piezoelectric mechanism of Sb and Bi monolayer materials and potential applications in nano-devices using their anomalous electromechanical response.

## ACKNOWLEDGMENTS

The authors gratefully acknowledge the support of NSFC (Grant No.11974269), the support by the Key Research and Development Program of Shaanxi (Grant No. 2023-YBGY-480), and the 111 projects 2.0 (Grant No. BP0618008). J.Z.L. acknowledges the support from ARC discovery projects (DP180101744) and HPC from National Computational Infrastructure from Australia. The authors also thank F. Yang and X. D. Zhang at the Network Information Center of Xi'an Jiaotong University for supporting the HPC platform. This work is also supported by State Key Laboratory for Mechanical Behavior of Materials.

**Figure 1**

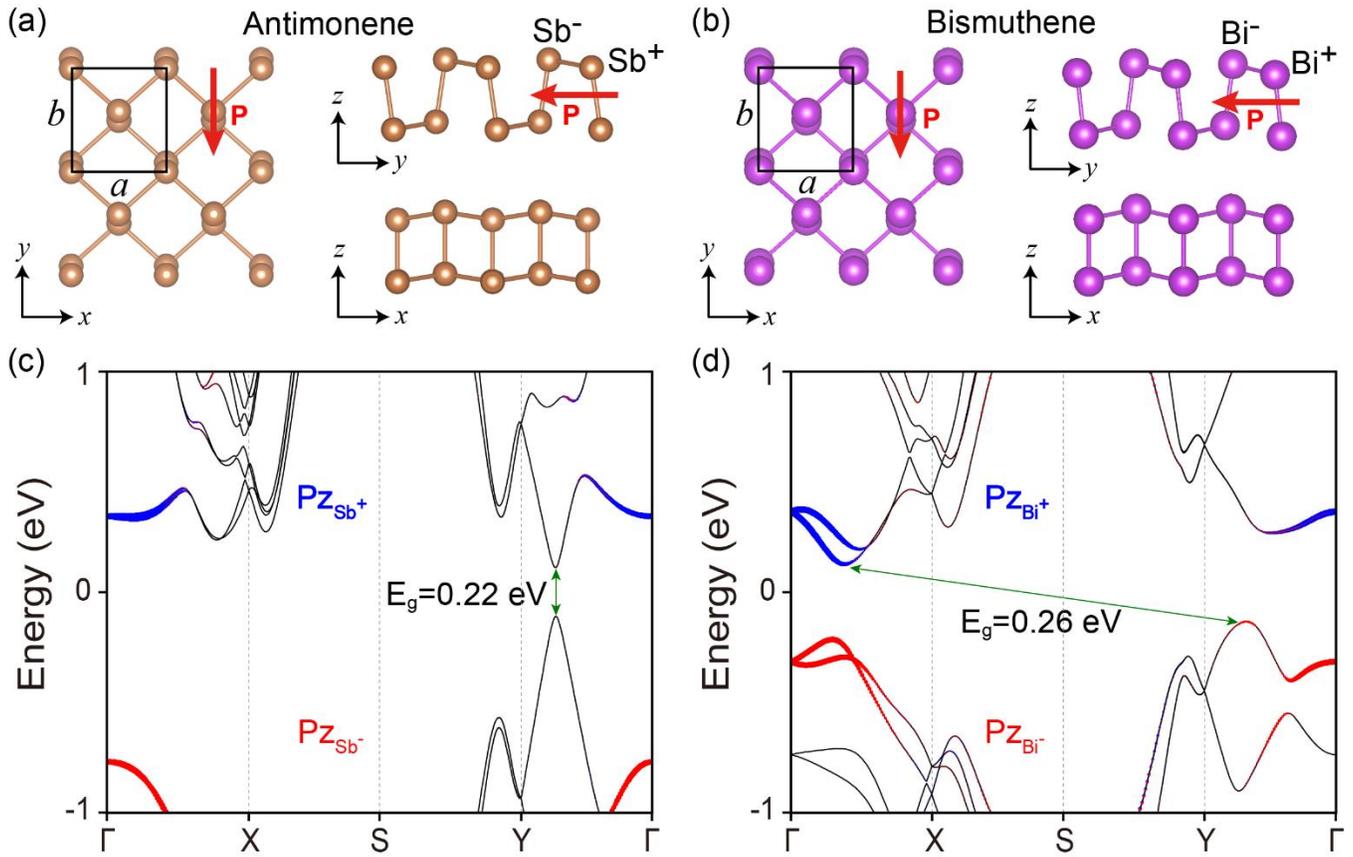

**FIG 1 Asymmetric structure of Sb and Bi monolayers.** (a)-(b) The atomic structure of Sb and Bi monolayers. Solid line boxes represent the unit cell of Sb and Bi. The red arrows indicate the intrinsic polarization direction. (c)-(d) Band structure of Sb and Bi monolayers. The size of red (blue) circles represents the contributions of the $p_z$ orbitals of Sb$^-$ (Sb$^+$) and Bi$^-$ (Bi$^+$). The valance band maximum (VBM) and conduction band minimum (CBM) were identified as the end of green arrows. The energy difference of them was the band gap ($E_g$).

**Figure 2**

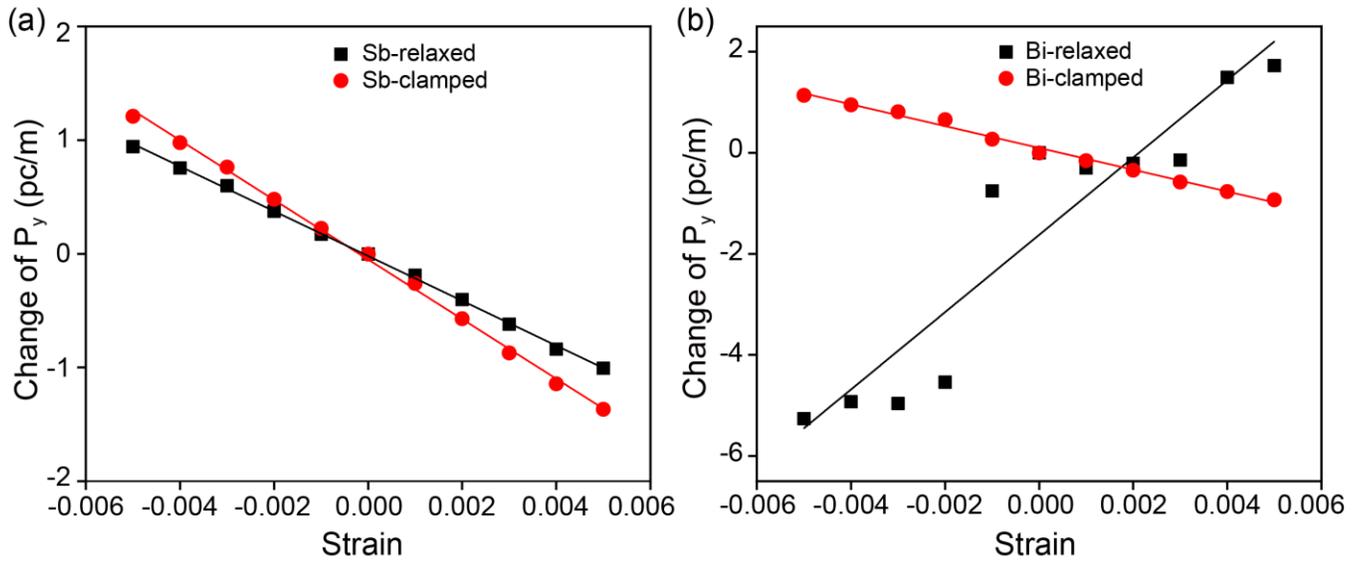

**FIG 2 The polarization of Sb (a) and Bi (b) as a function of uniaxial strain.** "Strain = 0" represent the ground states of Sb and Bi monolayers. The slopes of Sb/Bi-relaxed curves are the total piezoelectric coefficient ($e_{22}$) of Sb/Bi monolayers. The slopes of Sb/Bi-clamped curves are the clamped-ion terms ($e_{22}^0$) of Sb/Bi monolayers.

**Figure 3**

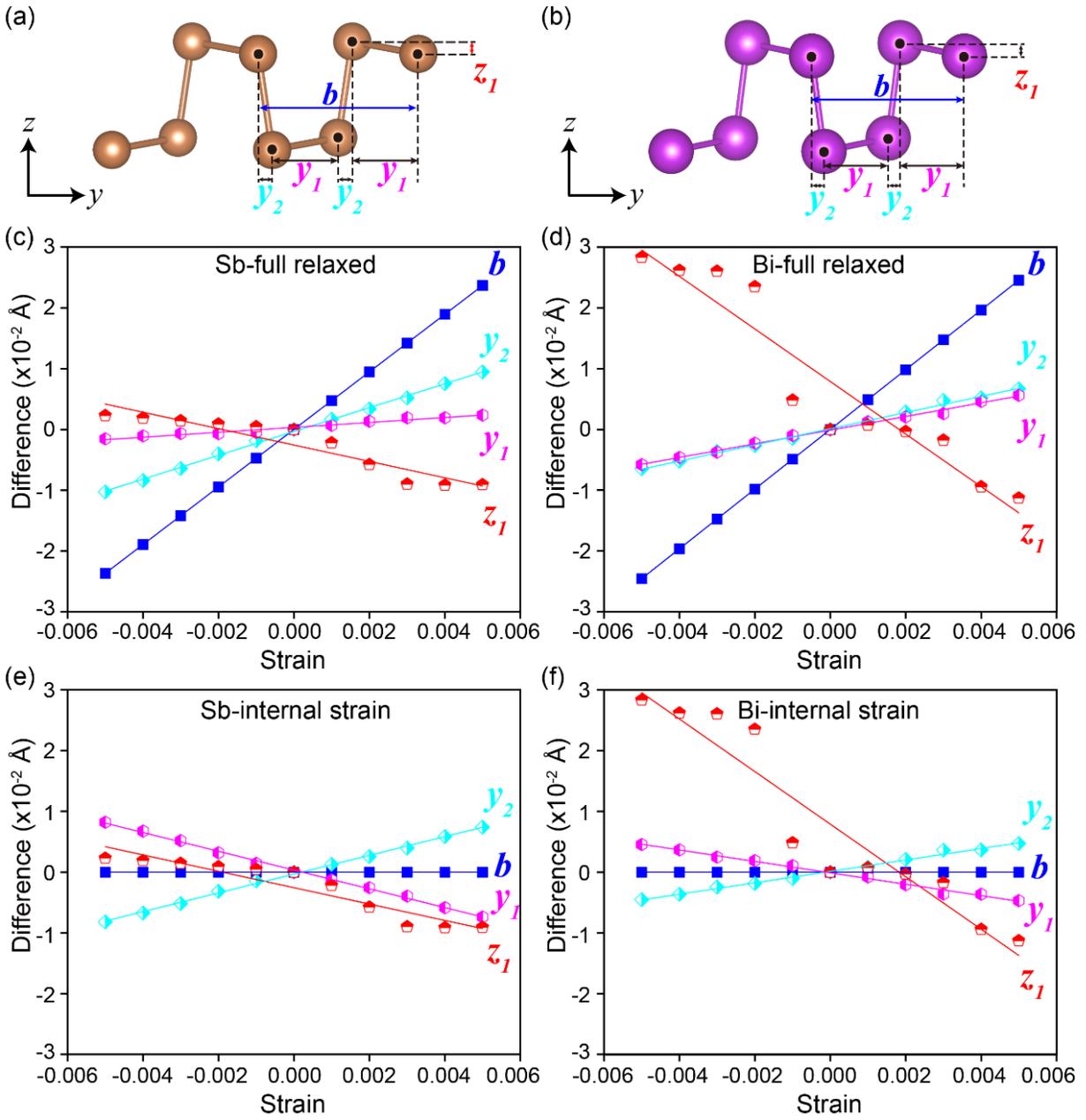

**FIG 3 Changes of structural parameters of Sb and Bi under uniaxial strain**. (a)-(b) Atomic structure of Sb and Bi. The parameter $b$ is the lattice constant in $y$ direction. $y_1$ and $y_2$ represent distances between adjacent atoms in $y$ direction, and the sum of $y_1$ and $y_2$ is equal to half the lattice constant $b$ ($y_1 + y_2 = 0.5b$). The parameter $z_1$ represents the distance between Sb$^-$(Bi$^-$) and Sb$^+$(Bi$^+$) atoms in $z$ direction. (c)-(d) Changes of full relaxed structural parameters of Sb and Bi. (e)-(f) Changes of structural parameters of Sb and Bi during the internal-strain process.

**Figure 4**

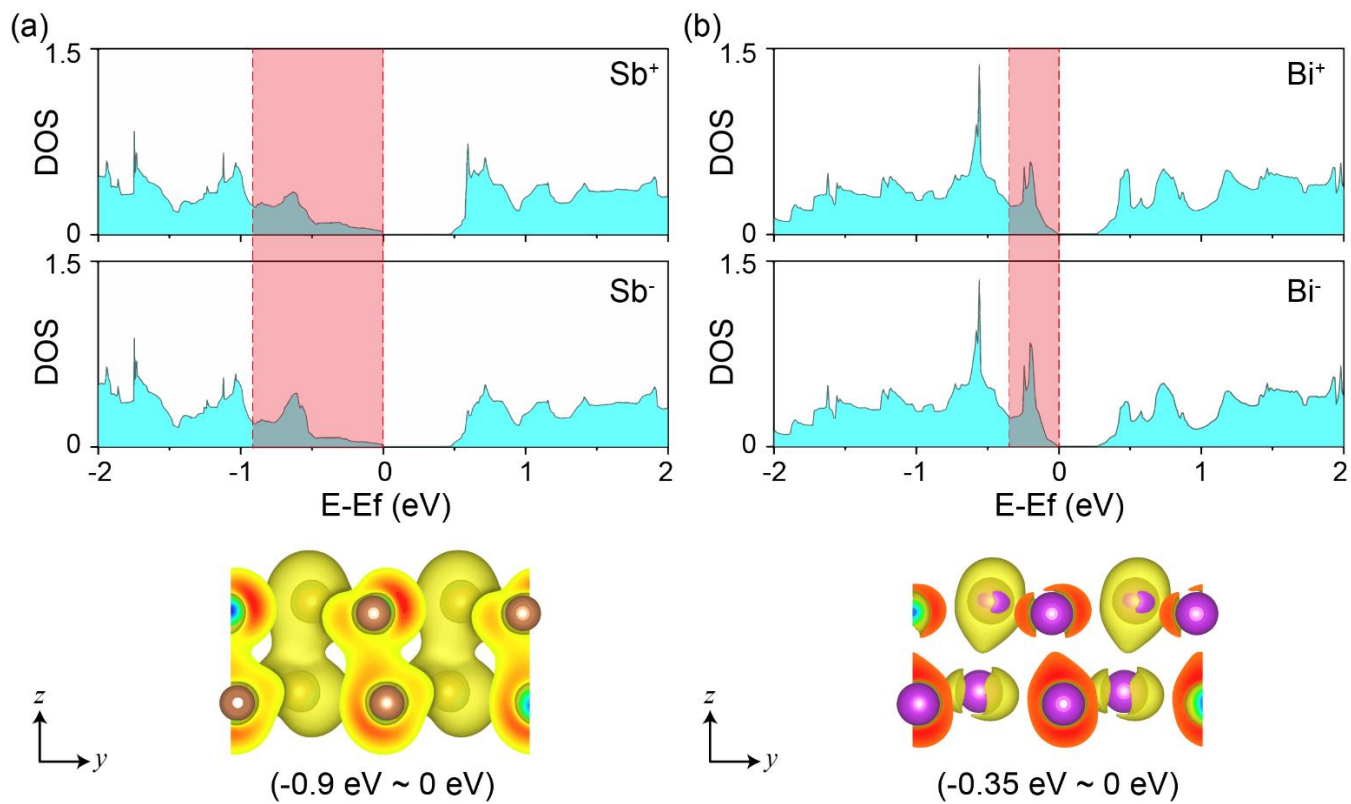

**FIG 4 Bonding analysis of Sb (a) and Bi (b) monolayers.** The density of states of $Sb^+$ ($Bi^+$) and $Sb^-$ ($Bi^-$) versus $E - E_f$ are shown in light blue, where $E_f$ denotes Fermi level. The red boxes highlight the difference in DOS between the two atoms near the valence band maximum, and the electron-density distribution in this energy range are visualized in the bottom sub-figures.

TABLE I The piezoelectric tensor elements $e_{22}$ of Sb and Bi monolayers. The $e_{22}$ is decomposed into the "clamped-ion" term ($e_{22}^0$) and the "internal-strain" term ($e_{22}'$).

|    | $e_{22}^0$ ($\times 10^{-10} c/m$) | $e_{22}'$ ($\times 10^{-10} c/m$) | $e_{22}$ ($\times 10^{-10} c/m$) |
|----|----|----|----|
| Sb | -2.623 | 0.652 | -1.971 |
| Bi | -2.157 | 9.810 | 7.653 |

TABLE II The Born effective charges of Sb and Bi monolayers (the ground state). Sb⁻ (Bi⁻) and Sb⁺ (Bi⁺) are two states of 2D Sb (Bi). By taking symmetry into account, the values were averaged and normalized.

|    | $Z_{12}^*$ (e) | $Z_{22}^*$ (e) | $Z_{32}^*$ (e) |
|----|----|----|----|
| Sb⁻ | -0.10 | -0.34 | -0.29/0.29 |
| Sb⁺ | 0.10 | 0.34 | -0.84/0.84 |
| Bi⁻ | 0.09 | -0.74 | -0.07/0.07 |
| Bi⁺ | -0.09 | 0.74 | -3.28/3.28 |

# Supplemental Information

**Unexpected Reversed Piezoelectric Response in Elemental Sb and Bi Monolayers**

Yunfei Hong[1], Junkai Deng[1,*], Qi Kong[1], Xiangdong Ding[1], Jun Sun[1], and Jefferson Zhe Liu[2,*]

[1]State Key Laboratory for Mechanical Behavior of Materials, Xi'an Jiaotong University, Xi'an 710049, China

[2]Department of Mechanical Engineering, The University of Melbourne, Parkville, VIC 3010, Australia

[*]Corresponding author

E-mail: junkai.deng@mail.xjtu.edu.cn; zhe.liu@unimelb.edu.au

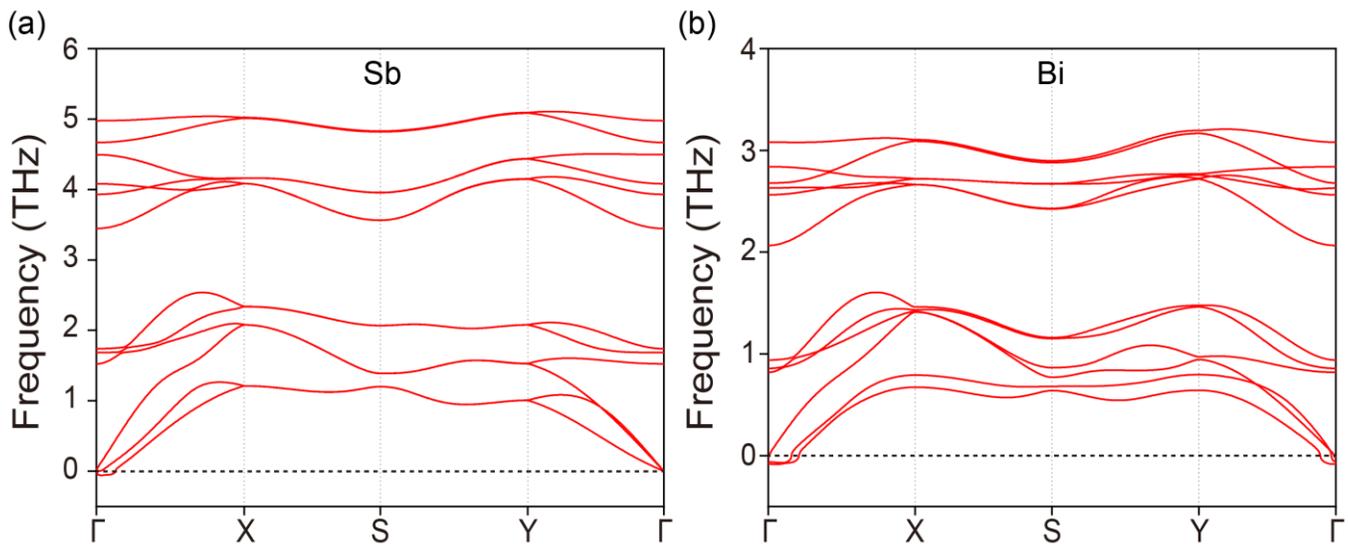

FIG S1: Phonon spectrum of Sb (a) and Bi (b) monolayers.

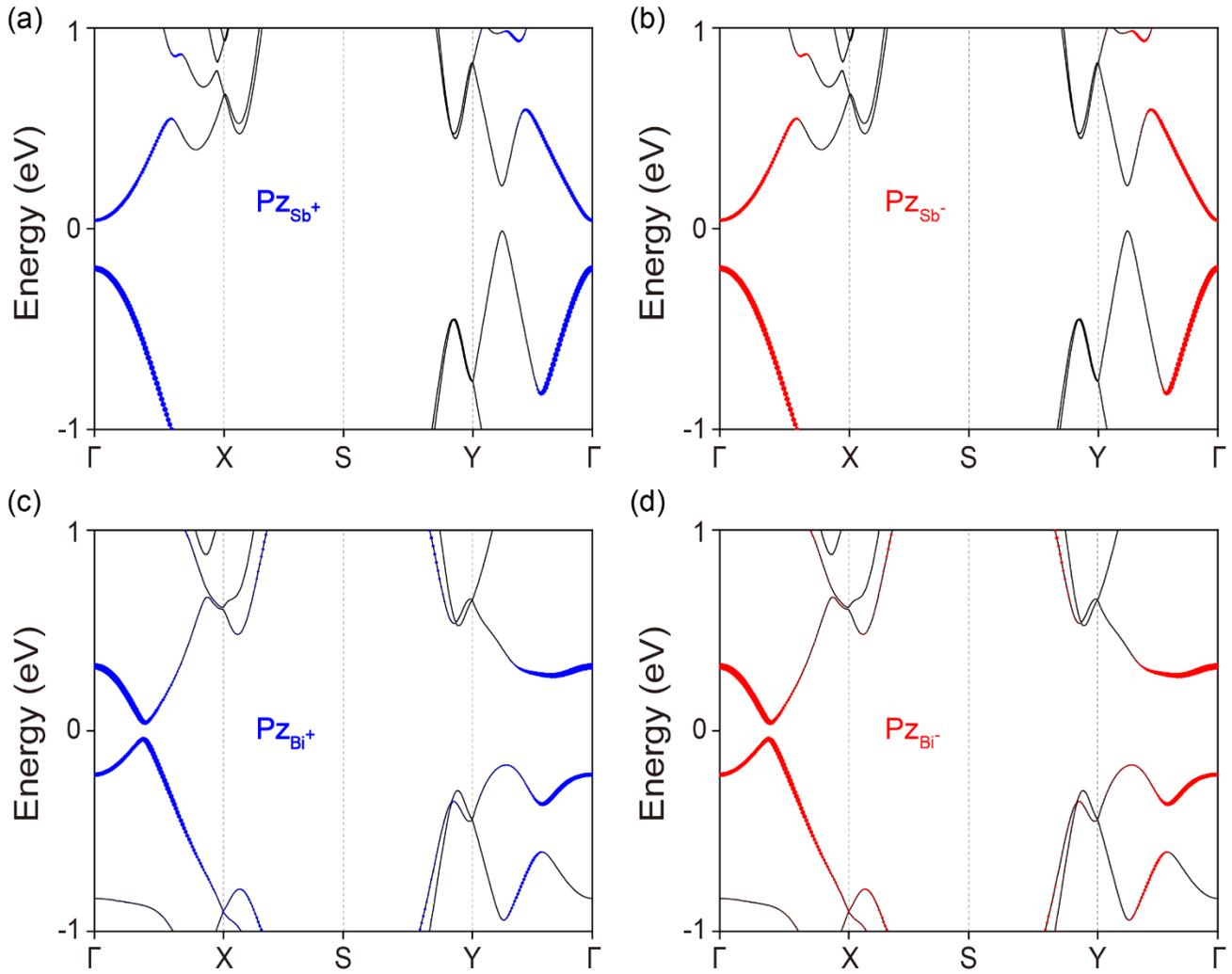

FIG S2: Band structure of Sb (a-b) and Bi (c-d) monolayers. The size of red (blue) circles represents the contributions of the $p_z$ orbitals of Sb⁻ (Sb⁺) and Bi⁻ (Bi⁺). The results revealed that there is no difference between the contributions of the $p_z$ orbitals of Sb⁻ (Bi⁻) and Sb⁺ (Bi⁺).

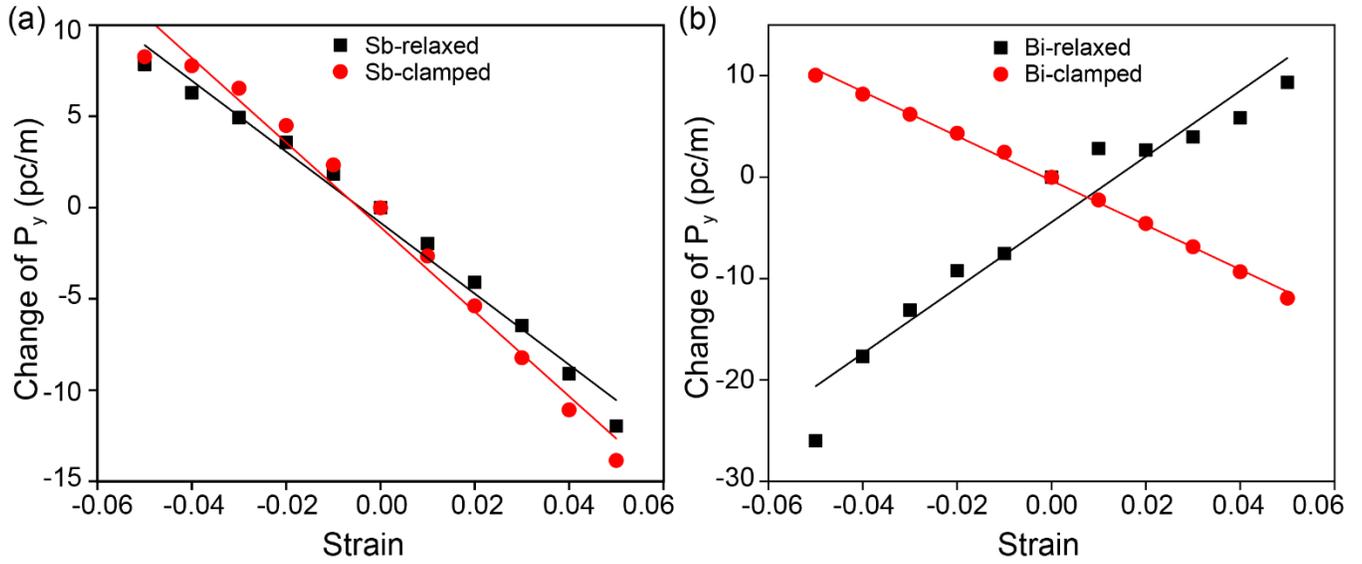

FIG S3: The polarization of Sb (a) and Bi (b) as a function of uniaxial strain. "Strain = 0" represent the ground states of Sb and Bi monolayers. The slopes of Sb/Bi-relaxed curves are the total piezoelectric coefficient ($e_{22}$) of Sb/Bi monolayers. The slopes of Sb/Bi-clamped curves are the clamped-ion terms ($e_{22}^0$) of Sb/Bi monolayers.

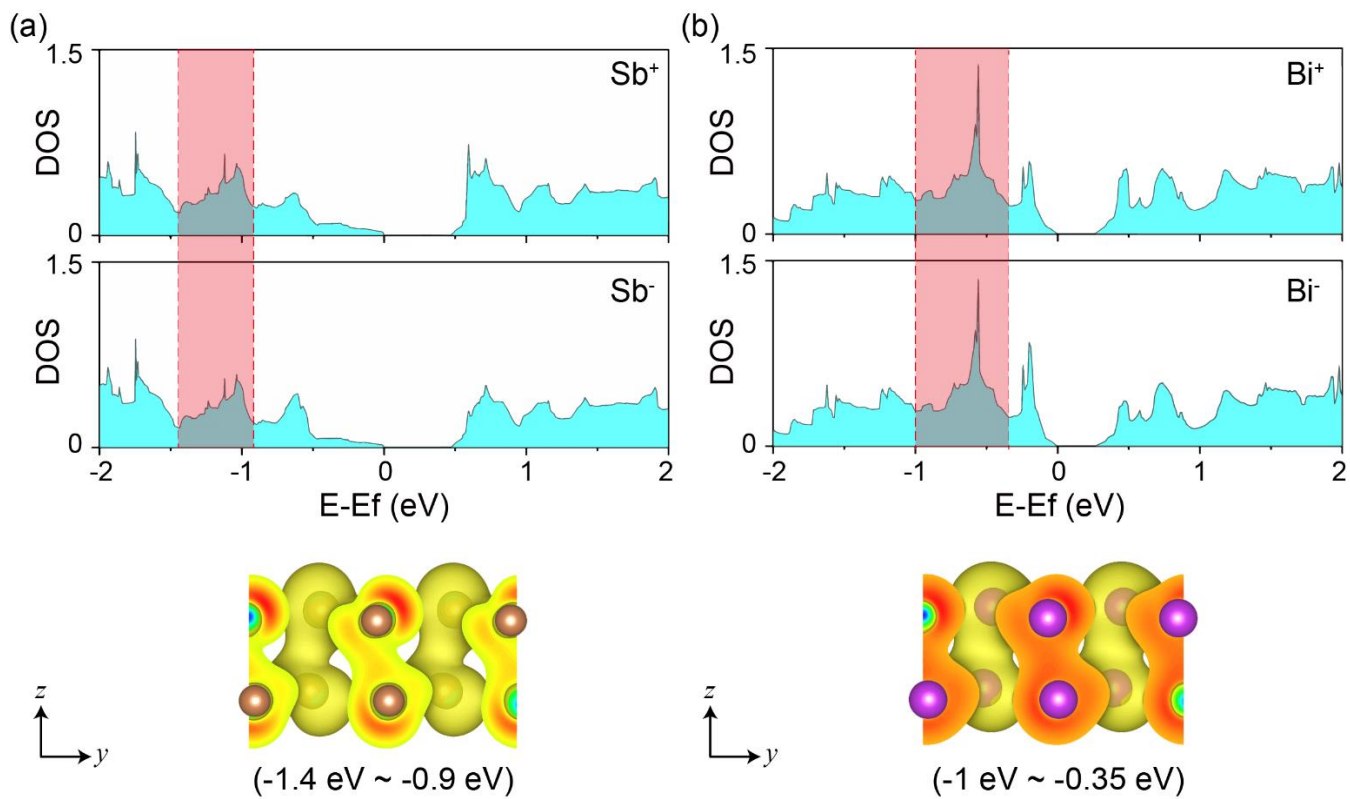

FIG S4: Bonding analysis of Sb (a) and Bi (b) monolayers. The density of states of $Sb^+$ ($Bi^+$) and $Sb^-$ ($Bi^-$) are shown in light bule. The red boxes highlight the difference in DOS between the two atoms, and the electron-density distribution in this energy range are visualized in the bottom sub-figures.

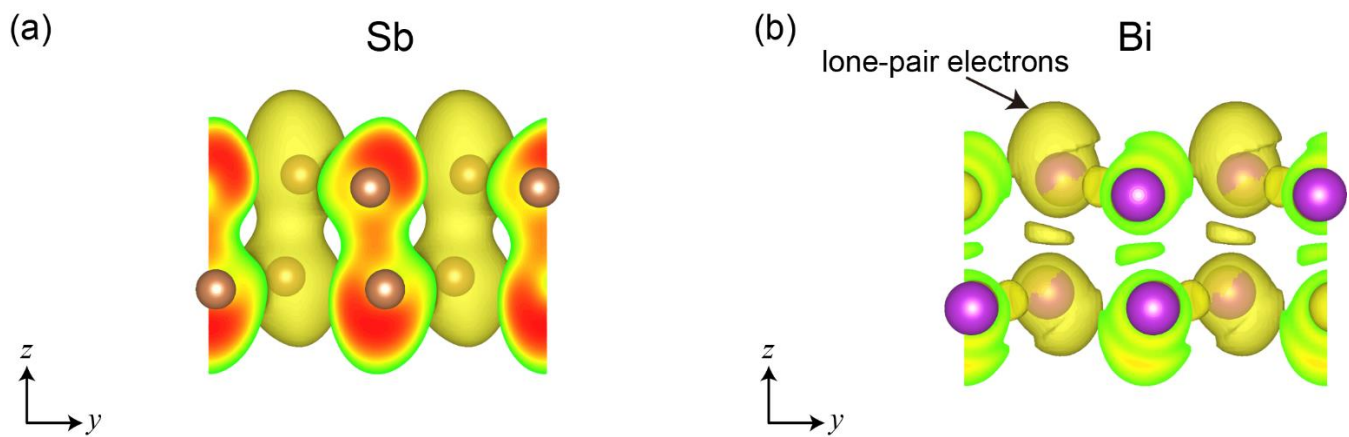

FIG S5: Electronic local function (ELF) analysis of Sb (a) and Bi (b) monolayers. The lone-pair electrons were indicated by the black arrow.